\documentclass[aps,
prl, superscriptaddress,
reprint]{revtex4-2}

\usepackage{natbib}
\setcitestyle{numbers}
\usepackage{graphicx}
\usepackage{amssymb, amsmath}
\usepackage{tabularx}
\usepackage{array}
\usepackage{multirow, makecell}

\usepackage[normalem]{ulem}

\usepackage{xcolor}

\begin{document}

\title{Incommensurable matter-wave jets in quasi-1D geometry}

\author{Tadej Me\v{z}nar\v{s}i\v{c}}
\email[]{tadej.meznarsic@ijs.si}
\affiliation{Jo\v{z}ef Stefan Institute, Jamova 39, SI-1000 Ljubljana, Slovenia}
\affiliation{Faculty of mathematics and physics, University of Ljubljana, Jadranska 19, SI-1000 Ljubljana, Slovenia}
\author{Rok \v{Z}itko}
\affiliation{Jo\v{z}ef Stefan Institute, Jamova 39, SI-1000 Ljubljana, Slovenia}
\affiliation{Faculty of mathematics and physics, University of Ljubljana, Jadranska 19, SI-1000 Ljubljana, Slovenia}
\author{Katja Gosar}
\affiliation{Jo\v{z}ef Stefan Institute, Jamova 39, SI-1000 Ljubljana, Slovenia}
\affiliation{Faculty of mathematics and physics, University of Ljubljana, Jadranska 19, SI-1000 Ljubljana, Slovenia}
\author{Jure Pirman}
\affiliation{Jo\v{z}ef Stefan Institute, Jamova 39, SI-1000 Ljubljana, Slovenia}
\affiliation{Faculty of mathematics and physics, University of Ljubljana, Jadranska 19, SI-1000 Ljubljana, Slovenia}
\author{Katja Arh}
\affiliation{Jo\v{z}ef Stefan Institute, Jamova 39, SI-1000 Ljubljana, Slovenia}
\affiliation{Faculty of mathematics and physics, University of Ljubljana, Jadranska 19, SI-1000 Ljubljana, Slovenia}
\author{Matev\v{z} Jug}
\affiliation{Jo\v{z}ef Stefan Institute, Jamova 39, SI-1000 Ljubljana, Slovenia}
\affiliation{Faculty of mathematics and physics, University of Ljubljana, Jadranska 19, SI-1000 Ljubljana, Slovenia}
\author{Erik Zupani\v{c}}
\affiliation{Jo\v{z}ef Stefan Institute, Jamova 39, SI-1000 Ljubljana, Slovenia}
\affiliation{Faculty of natural sciences and engineering, University of Ljubljana, A\v{s}ker\v{c}eva 12, SI-1000 Ljubljana, Slovenia}
\author{Peter Jegli\v{c}}\email[]{peter.jeglic@ijs.si}
\affiliation{Jo\v{z}ef Stefan Institute, Jamova 39, SI-1000 Ljubljana, Slovenia}
\affiliation{Faculty of mathematics and physics, University of Ljubljana, Jadranska 19, SI-1000 Ljubljana, Slovenia}

\date{\today}
\renewcommand{\arraystretch}{1.5}
\renewcommand\theadfont{\normalfont\bfseries}

\begin{abstract}

Matter-wave jets are ejected from a Bose-Einstein condensate subjected to a modulation of the interaction strength. For sufficiently strong modulation additional higher harmonic matter-wave jets emerge.
Here we report the first experimental observation of incommensurable ``golden'' $\frac{1+\sqrt{5}}{2}$ matter-wave jets in a Bose-Einstein condensate exposed to a single frequency interaction modulation. We study the formation of higher-order jets and the corresponding incommensurable density waves in quasi-one-dimensional geometry with numerical one dimensional (1D) Gross-Pitaevskii equation simulation.
We explore the process of jet formation experimentally and theoretically for a wide range of modulation amplitudes and frequencies and establish a phase diagram delineating different regimes of jet formation.
\end{abstract}

\maketitle

Incommensurability is encountered in aperiodic crystals with incommensurable phases \cite{Bak1982, Lifshitz2007, Janssen2012}, charge and spin structures \cite{Feng2013, Keimer2015, Wang2018, Miao2019, Storeck2020}, twisted moir{\' e} bilayer experiments \cite{Carr2020} with Hofstadter butterfly energy structure \cite{Lu2021}, and surface layers \cite{Hu2022}.
In the context of cold atoms, incommensurable optical lattices have been used to study wavefunction localization phenomena in disordered potentials within the Aubry-Andr\'{e} model \cite{Lye2007, Roati2008,Fallani2008,  Bordia2017, Schreiber2015, DErrico2014, Gadway2011, Lin2019, Reeves2014}.

Matter-wave jets occur when a BEC is subjected to a modulation of interaction \cite{Clark2017}. The modulation leads to the formation of density waves inside the BEC followed by the emission of jets \cite{Fu2018}. 
The microscopic process behind jet emission is photon-stimulated two-atom collisions \cite{Clark2017, Feng2019}, where a quantum of the magnetic field modulation energy is transferred to the atom pair. First-order jets form from pairs of atoms that each absorb half of the energy of the photon. 
Further collisions lead to higher-order jets having velocities that can either be rational multiples of first-order jet velocity (commensurate jets) or not (incommensurate jets).

In our previous work we studied the formation of first- and second-order commensurate jets in a quasi-one-dimensional geometry \cite{Meznarsic2020}. We observed the jets experimentally and in numerical simulations, however, no incommensurate jets were seen, because they only occur at higher modulation amplitudes.
Previous studies of matter-wave jets in two dimensions have revealed the formation of density waves inside the condensate prior to the emission of jets and explained the process of formation of higher-order jets \cite{Fu2018, Feng2019}. In Ref. \cite{Feng2019} higher-order jets with velocities $\sqrt{2}$ and $2$ times the velocity of first-order jets were directly observed, commensurate with the first order. Additionally, a hint of incommensurate $\sqrt{3}$ jets was reported.
It is important to note that in 2D geometry the $\sqrt{2}$ jets actually result from a density modulation with the velocity vector at a 45° angle with respect to the primary density modulation, so that each component of the velocity vector is, in fact, commensurate with the primary modulation. In contrast, $\sqrt{2}$ jets in 1D geometry result from a genuine incommensurability in the BEC. 

Here, we experimentally and numerically demonstrate the formation of incommensurable ``golden'' jets and other higher-order matter-wave jets in a quasi-1D BEC under a single-frequency interaction modulation. The incommensurable golden jets are precursors of the supercontinuum formation, which remains to be experimentally confirmed.

\begin{figure}[!b]
\centering
\includegraphics[width=\linewidth]{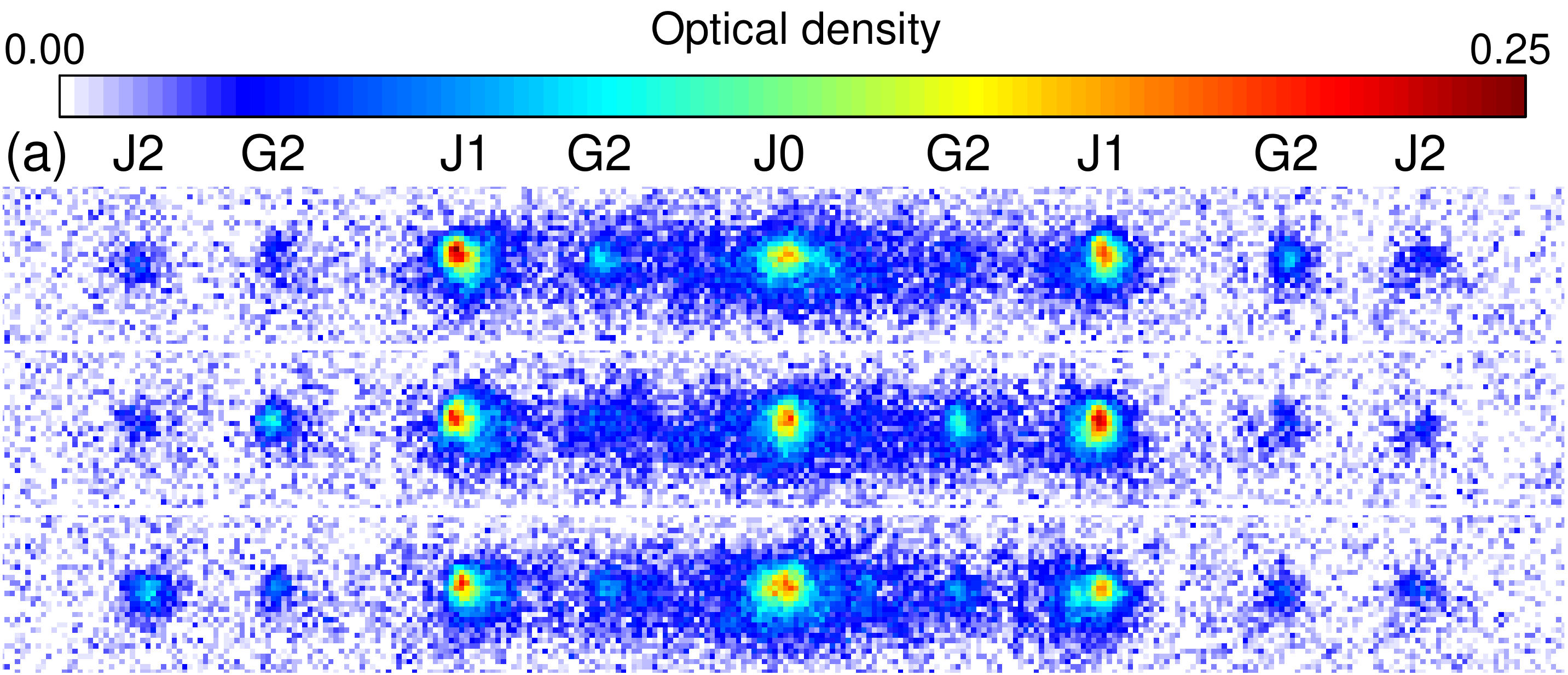}
\includegraphics[width=\linewidth]{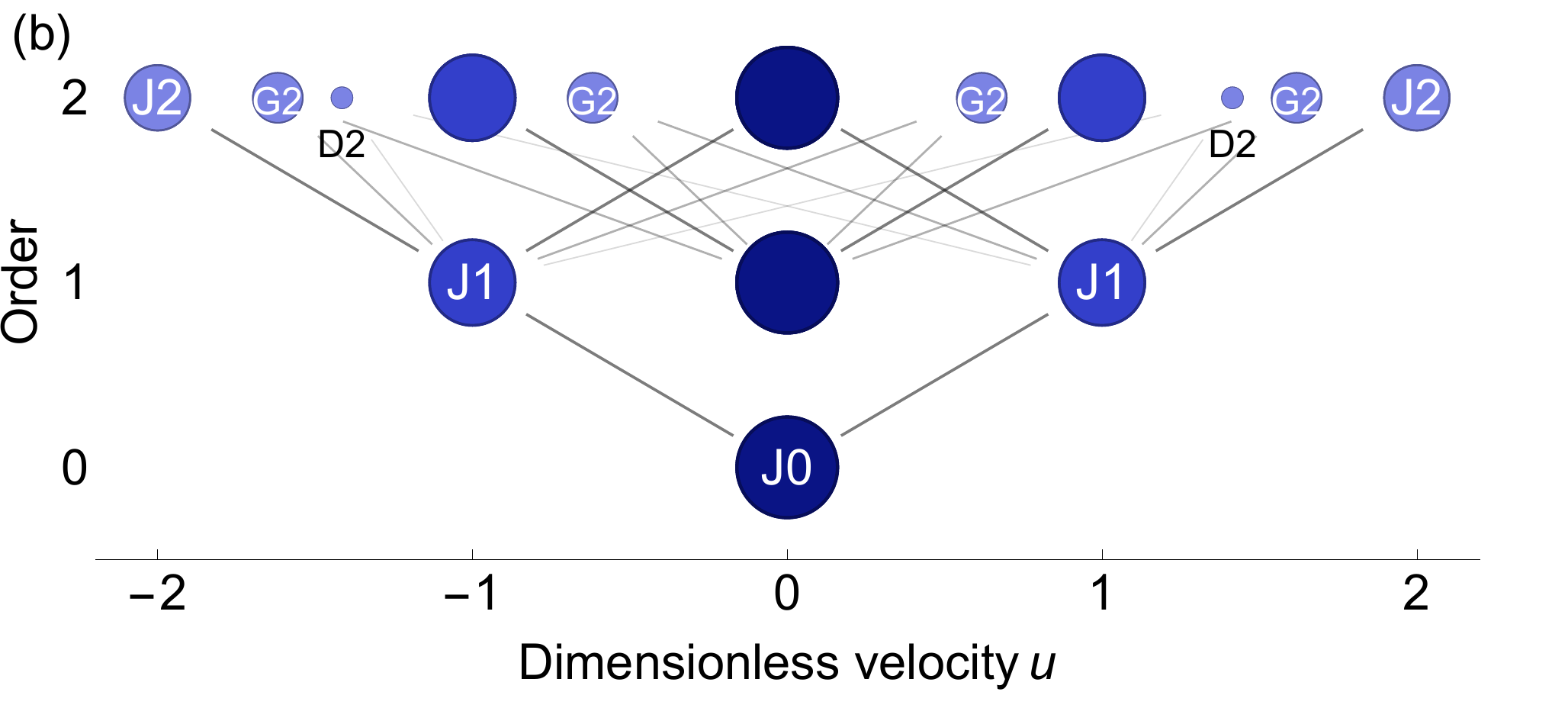}
\caption{\label{fig:1} 
(a) Absorption images of three different realizations of golden jets (modulation frequency 4~kHz, amplitude $a_\mathrm{ac} = 90a_\mathrm{0}$ ($k_\mathrm{ac} = 21.9$), $t_\mathrm{p} = 10$~ms, and $t_\mathrm{e} =40$~ms), showing the original BEC (J0), first order jets (J1), second order jets (J2) and revealing two pairs of ``golden'' jets (G2) (but not $\sqrt{2}$ (D2) jets).
(b) Schematic of possible jets for increasing number of subsequent collisions (order, encoded by color).
}
\end{figure}

\section{Results}
\subsection{Formation of matter-wave jets}
We start by preparing an almost pure BEC of approximately 4000 cesium atoms in a crossed dipole trap.  We release it into a channel with radial frequency $\omega_\mathrm{r} =  2\pi\cdot 90$~Hz and simultaneously change the interaction between the atoms to a value of $a_\mathrm{dc}$ that is close to zero via the wide $s$-wave Feshbach resonance \cite{Meznarsic2020} with zero-crossing at 17.1~G. This results in the formation of a BEC soliton. For 4000 atoms the BEC forms a soliton at interaction $a_\mathrm{dc} = -2.7a_0$ \cite{Meznarsic2019}, where $a_0$ is the Bohr radius.
To induce the emission of matter-wave jets, we modulate the magnetic field and therefore the interaction between the atoms as $a(t) = a_\mathrm{dc} + a_\mathrm{ac} \sin (\omega t)$ for a time $t_\mathrm{p}$, where $a_\mathrm{ac}$ and $\omega$ are the amplitude and the frequency of the modulation, respectively  \cite{Clark2017} (see Methods).
We let the system evolve for $t_\mathrm{e}$ in the channel and take an absorption picture after a 15~ms time-of-flight expansion. The results are shown in Fig.~\ref{fig:1}(a).
In addition to previously observed jets of first order (J1) and second order (J2) \cite{Meznarsic2020}, we observe two pairs of jets with dimensionless velocities 
\begin{align}
\label{eq:1}
u_\mathrm{G2} = \frac{v_\mathrm{G2}}{v_\mathrm{0}} =  \pm\frac{1\pm\sqrt{5}}{2},
\end{align}
where $v_\mathrm{0} = \sqrt{\hbar \omega / m}$ is the velocity of the first-order jets (J1), $\hbar$ is the reduced Planck constant, and $m$ is the mass of an atom. 
We name these ``golden jets'' (G2) for their golden ratio dimensionless velocities.
The irrational velocity ratio between first order and golden jets implies the coexistence of two incommensurable density waves inside the condensate prior to jet emission.
Fig.~\ref{fig:1}(a) shows three experimental realizations showing golden jets. As can be seen in Fig.~\ref{fig:1}(a) the golden jets are not necessarily symmetric around J0 and change in each experimental run despite the supposed symmetry of the involved formation processes.
We attribute this to the asymmetry in the initial BEC which is slightly inhomogeneous and fluctuates between experimental runs.

\begin{table}[t]
\caption{\label{table:1} Jet orders, abbreviations, and formation processes. $u_{1, 2}$ are the dimensionless velocities of the initial atom pair and $u_{1, 2}'$ are the dimensionless velocities of the resulting jets.
Beyond $2^\mathrm{nd}$ order, only jets observed in the simulations are included. }
\begin{center} 	
\begin{tabular}{ c c c }
 \hline
 \hline
 \multirow{2}{*}{\thead{Order}} &\multirow{2}{*}{\thead{Abbrev.}} & \thead{Formation process} \\ [-1ex]
 & &  $(u_{1},u_{2})\rightarrow (u^\prime_{1},u^\prime_{2})$ \\ \hline
 $0^\mathrm{th}$ & J0 & / \\ \hline
 $1^\mathrm{st}$ & J1 & $(0,0)\rightarrow (-1,1)$ \\ \hline
 \multirow{3}{*}{$2^\mathrm{nd}$} & J2 & $\pm (1,1)\rightarrow (0,\pm 2)$ \\
 	& G2 & $(0,\pm 1)\rightarrow \pm \left(\frac{1-\sqrt{5}}{2}, \frac{1+\sqrt{5}}{2}\right)$ \\
 	& D2 & $(-1,1)\rightarrow \left(-\sqrt{2},\sqrt{2}\right)$ \\ \hline
 \multirow{2}{*}{$3^\mathrm{rd}$} & J3 & $\pm (2,2)\rightarrow \pm (1,3)$ \\	
 	& G3 & $\pm (1, 2)\rightarrow \pm \left(\frac{3-\sqrt{5}}{2}, \frac{3+\sqrt{5}}{2}\right)$ \\
 \hline
 \hline
\end{tabular}
\end{center}
\end{table}

The origin of golden jets can be understood from the conservation of energy
$\frac{mv_1^2}{2} + \frac{mv_2^2}{2} + \hbar\omega = \frac{mv_1^{\prime 2}}{2} + \frac{mv_2^{\prime 2}}{2}$, and momentum: $mv_1 + mv_2 = mv_1^{\prime} + mv_2^{\prime}$,
where $v_i$ ($v_i^{\prime}$) are the initial (final) velocities of the atoms and the momentum of the photon is neglected.
A pair of atoms in J1 is formed from a pair of colliding atoms from the original condensate (J0) by absorbing one quantum of energy from the modulating magnetic field.
Similarly, a pair J0-J2 forms from a pair of J1 atoms with the same velocity and a pair G2-G2 from a pair J0-J1.
In general the final velocities of the resulting jets are 
\begin{align}
\label{eq:2}
u^\prime_{1,2} = \frac{(u_1 + u_2) \pm \sqrt{(u_1-u_2)^2 + 4}}{2},
\end{align}
where $u_{1,2}$ are the initial dimensionless velocities of the atom pair.
With each absorbed photon the number of possible jets increases.
By the number of photons that an atom in a jet has absorbed we define the order of the jets: the initial BEC is zeroth order, J1 are first order, J2, G2 and D2 are second order and so on, see Fig.~\ref{fig:1}(b) and Table~\ref{table:1}.
The  total number of jets increases as a double exponential with order, going as 1, 3, 11, 109, 11605,$\hdots$, and can be approximately calculated as $1.7949^{2^n}$ (from the fit up to $n=4$, where $n$ is the order). Experimentally, we observe J1, J2 and G2, but we have not been able to observe jets with dimensionless velocity $\pm \sqrt{2}$ (D2) or any third order jets. 
J2 are formed from J1 which move with the same velocity so the process is equivalent to the primary process in a moving frame, while the atoms that form G2 move relative to each other, which reduces the interaction time, making golden jets rare. For D2 the velocity difference is two times higher, explaining why we do not observe them.
Additionally, the formation of J2 is more probable due to bosonic stimulation, since they form in pair with J0, which is present in the system beforehand \cite{Feng2019}.

\begin{figure}
\centering
 \includegraphics[width=1\linewidth]{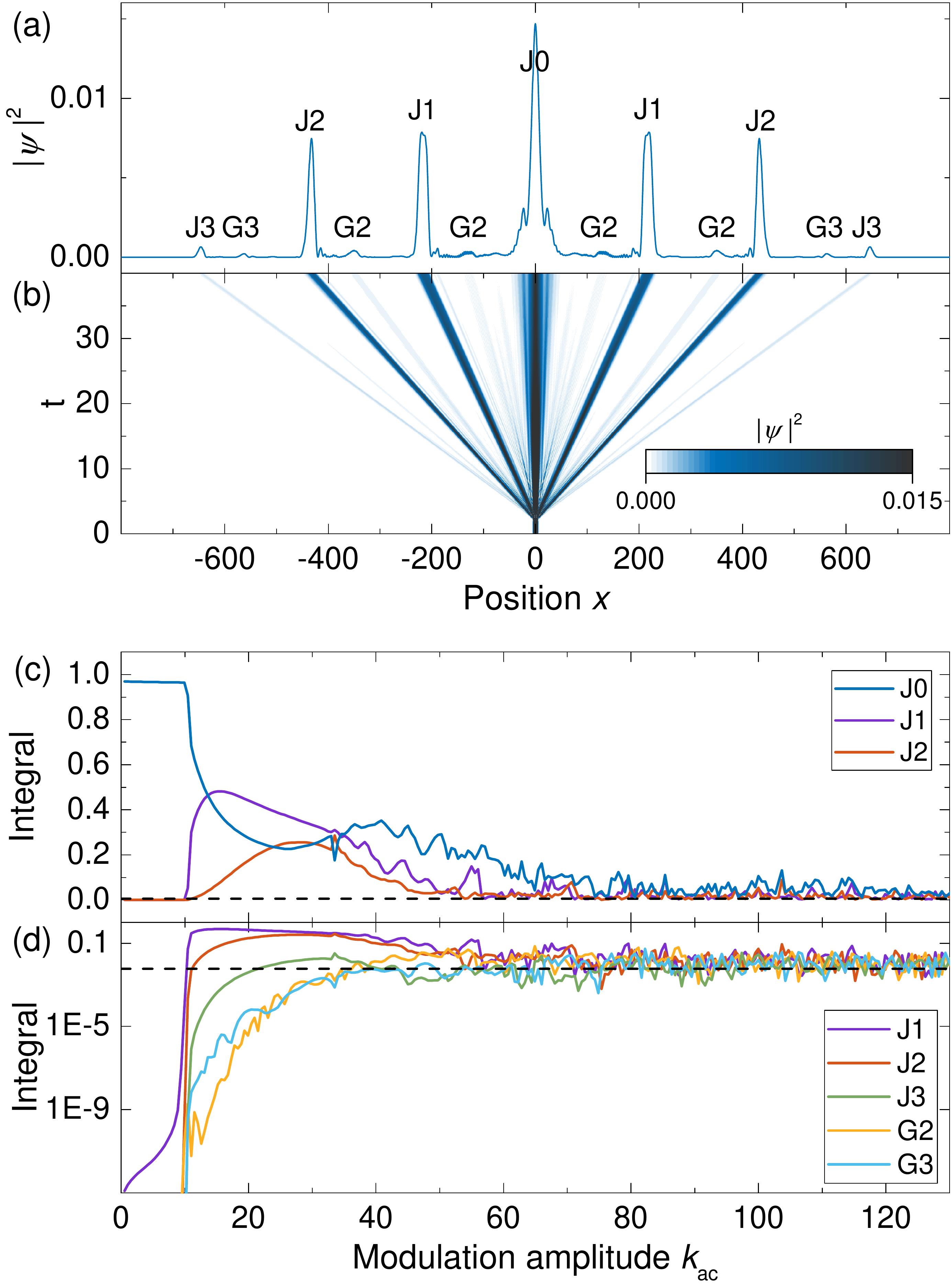}
\caption{\label{fig:2} 
(a) The final density profile ($t=40$) and 
(b) the time evolution of the density profile of the simulation for $\omega = 33$, $k_\mathrm{dc} = -0.56$, $k_\mathrm{ac} = 27.1$, $t_\mathrm{p} = 8\pi$ and the fraction of atoms in different jets in (c) linear and (d) logarithmic scale.
The horizontal line at 0.006 represents the threshold used to construct the phase diagram.
}
\end{figure}

\begin{figure*}[!t]
\centering
\includegraphics[width=\linewidth]{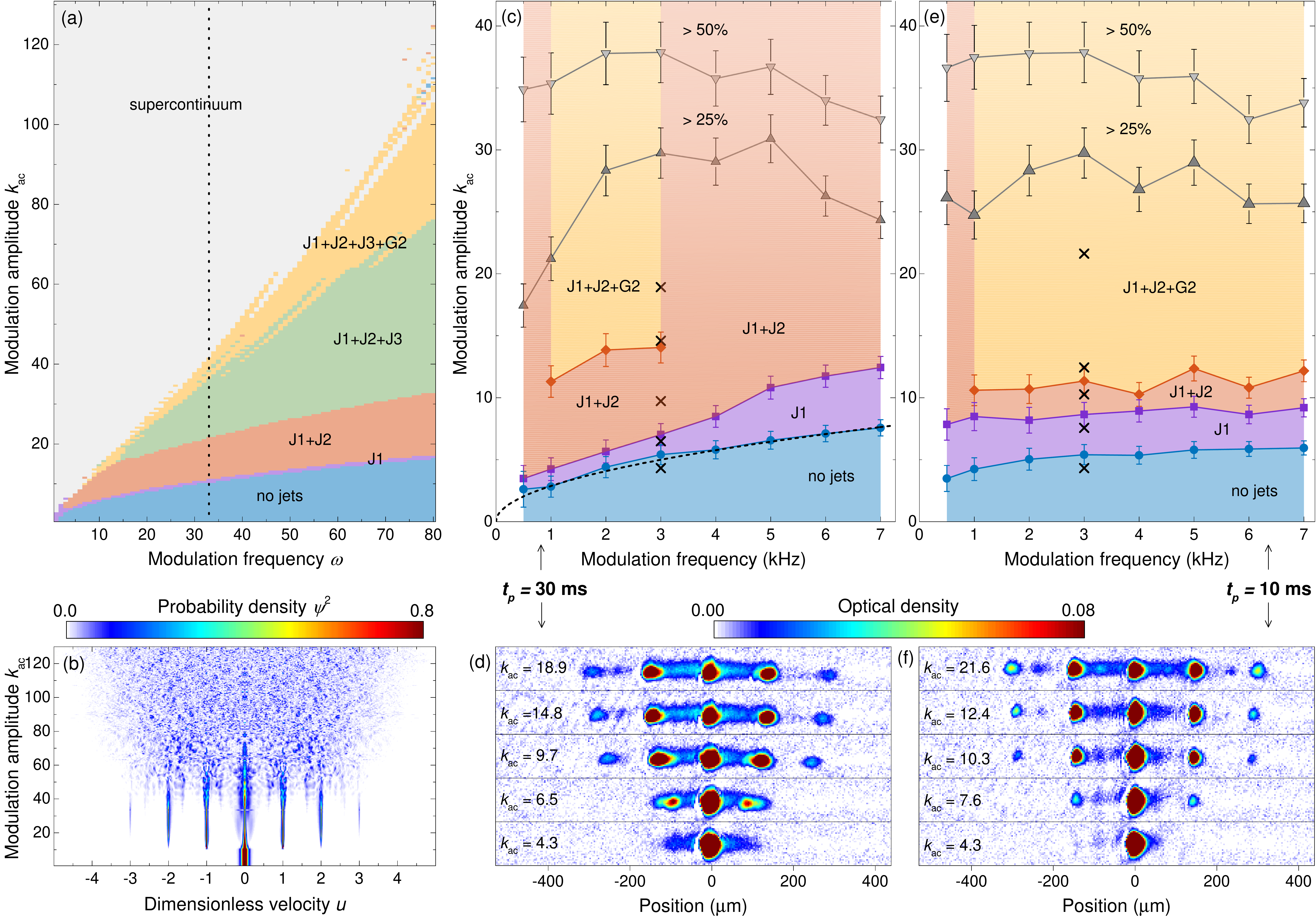}
\caption{\label{fig:3} 
(a) Phase diagram of jets from GPE simulations. Different colors represent regions where  different types of jets are above the threshold.
(b) Final density profiles in momentum space $|\psi(q)|^2$  ($t=8\pi$) for $\omega = 33$ [along the dotted line in (a)]. (c) and (e) Experimental phase diagram of jets, measured at $t_p = $ 30 ms and 10 ms. The grey lines indicate where the number of atoms falls by 25\% and 50\% from the initial number of atoms in the condensate. The dashed black line in (c) is a fit to the boundary between the "no jets" and "J1" region, proportional to $\omega^{1/2}$. The error bars correspond to the variance  of the atom number and the precision of determining  the amplitude corresponding to the boundary between the regions. (d) and (f) Averages of 50 absorption images as examples of different points in the phase diagram (marked with black crosses in the diagrams above). From bottom to top the images represent regions with: no jets, only J1, J1 and J2, and J1, J2 and G2 (top two images). The images were taken for evolution time $t_e = $ 42.7 ms. $k_\mathrm{ac} = Na_\mathrm{ac}/a_r$, where $N = 4000 \pm 200$ is the number of atoms in the BEC, before we start modulation. $a_\mathrm{ac}$ is calculated from the amplitude of the pulse by a calibration determined using the Feshbach resonance (see Methods).
}
\end{figure*}

\subsection{Numerical simulation of Gross-Pitaevskii equation}
We turn to numerical simulations of 1D Gross-Pitaevskii equation (GPE) to better understand the conditions for the formation of each order of jets.
As in our previous work \cite{Meznarsic2020}, we start with a soliton as the initial state and modulate the interaction parameter $k(t) = Na(t)/a_r = k_\mathrm{dc}+k_\mathrm{ac}\sin (\omega t)$ for a time $t_\mathrm{p} = 8\pi$ with different frequencies $\omega$ and amplitudes $k_\mathrm{ac}$.
We leave the wavefunction to evolve for $t_\mathrm{e}$ and examine the resulting density profile $n(x)=|\psi (x)|^2$. The details about the simulation are provided in Methods and in our previous work \cite{Meznarsic2020}.
In the simulations, the time $t$ is given in units $1/\omega_r$, $\omega$ in units $\omega_r$, and distances in units $a_r = \sqrt{\hbar/m\omega_r}$.
 
An example of the final density profile can be seen in Fig.~\ref{fig:2}(a) where all visible jets are marked. Fig.~\ref{fig:2}(b) shows the time evolution of the density profile from the start of modulation.
In contrast with the experiment, we observe additional jets of third order J3 and G3 which form from J2-J2 and J1-J2 pairs, respectively.
Surprisingly, G2 and G3 have comparable amplitude even though G2 jets are second order whereas G3 only arise in third order; this seems to be due to the large amplitude of J2 jets (from which G3 are formed) in the simulation.

To quantify the size of jets, we calculate the integrals over an interval around each peak in the density profile in momentum space $\vert \psi (q) \vert^2$, which gives the fraction of the atoms in jets of each order. For J0 we integrate from -0.56 to 0.56 and for other jets we integrate an area $\pm0.16$ around each jet.
The total integral is normalized, $\int |\psi(q)|^2 dq = 1$, and $\psi (q)$ is the Fourier transform of $\psi (x)$.
The fraction in the central peak J0 only starts decreasing after the threshold amplitude for J1 formation is reached, as can be seen in Fig.~\ref{fig:2}(c).
The J1 first increases steeply with amplitude and then levels out due to J0 depletion and exitations to the higher-order jets.
It can be seen that only the first order has a sharp threshold (Fig.~\ref{fig:2}(c)), while the higher orders increase smoothly from zero once the first order exists (Fig.~\ref{fig:2}(d)).
D2 are not observed even in simulations due to the onset of a wide background
at high amplitudes needed to create them (discussed below).

Based on calculated atom fractions, we construct a phase diagram of visibility of different sets of jets, shown in Fig.~\ref{fig:3}(a).
A new region is declared to exist when the fraction for the next jet exceeds 0.006.
We choose this value so that it is much larger than the noise in the simulation, but smaller than the atom fraction of the jets at high modulation amplitudes.
As the amplitude $k_\mathrm{ac}$ increases, additional jets appear, represented by new regions in the phase diagram.
As discussed above, the transitions are actually continuous and not discrete, except from the ``no jets'' to the ``J1'' region. G3 jets occur in parts of the region marked with "J1+J2+J3+G2" in Fig.~\ref{fig:3}(a).

Fig.~\ref{fig:3}(b) shows the final profiles for a fixed frequency $\omega = 33$.
The onset of J1, J2 and other jets can be clearly seen as the modulation amplitude increases.
For very high amplitudes the discrete jets start disappearing in the increasingly uniform background of a multitude of higher-order jets.
The observed spread from the single momentum mode to a wide momentum distribution resembles the generation of supercontinua in nonlinear optical media, where various instabilities cause an initially narrow frequency spectrum to disperse into a wide continuum \cite{Dudley2006}.
The boundary for the supercontinuum can be determined with an entropy-like measure $S = -\int \vert \psi(q) \vert^2 \log \left( \vert\psi(q) \vert^2 \right) dq$, where we integrate over the whole interval.
We scale the threshold with the square root of the modulation frequency, to take into account the scaling of the jet velocities with modulation frequency.
When $S> 0.38\sqrt{\omega}  $ we say that it has transitioned into the supercontinuum region (the transition is actually gradual, the threshold value was chosen to be above numerical noise and visually without separate jets). 
The threshold boundary for J1 formation is proportional to $\omega^{1/2}$, as previously reported \cite{Meznarsic2020, Clark2017}.

\subsection{Experimental phase diagram of jet formation}
We explore the phase diagram of jet formation experimentally as well. Fig. ~\ref{fig:3}(c) and (e) show the experimental phase diagrams of jet formation for modulation times $t_p =$ 10~ms and 30~ms, respectively. We determined the regions of the phase diagram where we observe J1, J1 and J2 (J1+J2), and J1, J2 and G2 (J1+J2+G2). 

In contrast to simulation, in the experiments, the number of atoms in jets varies from shot to shot; for example, see Fig.~\ref{fig:1}(a). 
This was previously observed in 2D \cite{Hu2019}, where the distribution of population in angular modes resembles the distribution of thermal radiation.
More information on the statistics of the number of atoms in each type of jet is available in the Supplemental Material (Fig.~S3). Because of this variance, gathering statistics is necessary to determine the presence of the phase at each point in the phase diagram. We choose the boundary to be at  the point where at least one image out of ten shows the corresponding jets. Examples of averages of 50 absorption images of jets from different parts of the phase diagram are shown in  Fig.~\ref{fig:3}(d) and (f) and in the Supplemental Material (Fig.~S2). Both Fig.~\ref{fig:3}(d) and (f) show experimental images for modulation frequency 3~kHz, but (f) is for the shorter modulation time $t_p=$ 10~ms and (d) is for $t_p=$ 30~ms. Note that in  Fig.~\ref{fig:3}(d) the distance between the jets and the initial condensate varies for different modulation amplitudes. This is because jets are emitted sooner if the modulation amplitude is higher (see Supplemental Material  Fig.~S6). This effect is not as pronounced Fig.~\ref{fig:3}(f) due to the shorter modulation time.

The range of frequencies in the experimental and the theoretical phase diagram are similar (the frequencies shown in Fig. ~\ref{fig:3}(a) range from 0 to 7.2 kHz (from 0 to 80 in units of $\omega_r = 2\pi \cdot 90 \mathrm{Hz}$), whereas the experimental phase diagrams in Fig. ~\ref{fig:3}(c) and (e) were measured for frequencies between 1 kHz and 7 kHz). On the other hand, the values of $k_\mathrm{ac}$ required for excitation of jets are approximately two times lower in the experiment than in simulation. The difference is most probably because the simulation describes a true 1D system, while in the experiment, the BEC is a 3D object with jets confined into quasi-1D system by a dipole trap beam.

For $t_p = $ 30 ms, we are not able to observe golden jets for modulation frequencies higher than 3 kHz, even though they are observable using the shorter $t_p =$ 10~ms. 
In the simulation, the modulation time is $t_p = 8\pi$, which corresponds to 44~ms. In the experiment, we used shorter modulation times, because longer modulation times cause more atom losses and increase the incoherent background, formed by the excited atoms that are not part of the jets \cite{Meznarsic2020}, making the occurrence of jets rarer and more difficult to observe (see Supplemental Material Fig. S1 and S2).  However, we see from the experimental phase diagrams that the shape of the threshold does depend on $t_p$. For a short $t_p$ there is no strong dependence on the modulation frequency, whereas for the longer $t_p$ the threshold boundary for the formation of jets is proportional to $\omega^{1/2}$, as expected from the simulation and previous experiments \cite{Meznarsic2020}. Additionally, the longer modulation times sometimes cause the jets to be emitted twice, making the interpretation of the absorption images less reliable. 

As we increase the amplitude of modulation in the experiment, the total number of atoms starts decreasing and the incoherent background between the jets becomes stronger. This can be seen in Fig.~\ref{fig:3}(c) and (e), where we plot the thresholds at which the total number of atoms is 25\% and 50\% smaller than the number of atoms in the initial BEC. The decrease in the number of atoms is likely due to partial collapses of the BEC, caused by the strong attractive interactions the BEC is subjected to during the modulation, leading to three-body losses. The incoherent background is formed by the excited atoms that are not part of the jets. Their velocities have random directions and that appears as a homogeneous background between J0 and J1, and to a lesser extent between J1 and J2. The momentum distribution of the incoherent background remains of the same width for increasing amplitude and does not increase beyond the position of J2 jets as the supercontinuum would.
Atom losses and increased fraction of incoherent background, both not accounted for in the simulation, decrease the effective modulation amplitude and prevent the supercontinuum from forming.

To summarize, the experimental phenomena that the 1D GPE simulation  does not account for are the atom losses into the incoherent background and three-body losses. For a complete description of the experiment, one would not only need to take into account the three-dimensionality of the experiment, but also the higher order interaction between atoms (e. g. three-particle interaction) and incoherent processes, that are not described by the mean field picture. Despite all of these differences, the discrepancy between the simulation and the experiment only becomes significant at higher modulation amplitudes.

\section{Discussion}
In this work, we show the formation of matter-wave jets with golden ratio dimensionless velocity $u_\mathrm{G2}={(1+\sqrt{5})/2}$, which  form when a BEC in a quasi-1D geometry is subjected to a single-frequency interaction modulation. These jet velocities and the corresponding density waves are not commensurate with the previously observed jets with dimensionless velocities $u_\mathrm{J1} = 1$ and $u_\mathrm{J2} = 2$.  A hint of incommensurate $\sqrt{3}$ jets has previously been observed in similar experiments with 2D geometry \cite{Feng2019}, but golden jets are the first type of incommensurable jets observed in 1D. It needs to be noted that the occurrence of incommensurate density waves can easily be achieved with two-frequency driving (like in Ref. \cite{Zhang2020}, but with the frequencies that produce first-order jets that are not commensurable, see Supplemental Material Fig. S4), but here incommensurability emerges from single-frequency driving.

Numerically and experimentally, we explore the formation of incommensurable jets for a wide range of modulation frequencies and amplitudes, resulting in a phase diagram of possible jets, while also uncovering an unexpected supercontinuum region. 
The supercontinuum likely arises due to the double-exponential growth of the number of possible modes, while the wavenumber only grows linearly, resulting in a densely populated momentum space.
The supercontinuum is potentially experimentally observable if the incoherent background  could be reduced by increasing the radial confinement and suppressing three-body losses.

The demonstrated formation of incommensurable matter-wave jets could, for example, be used in multispecies experiments. The incommensurable density waves inside one species could be used as a disordered potential for the other species (provided there exist appropriate Feshbach resonances). 
This would eliminate the need for an external disordered potential or a regular lattice that is usually needed to capture impurities, which act as a source of disorder  \cite{Lye2007, Roati2008,Fallani2008,  Bordia2017, Schreiber2015, DErrico2014, Gadway2011, Lin2019, Reeves2014}.

\section{Methods}

\subsection{Preparation of the BEC}

We start the preparation of the BEC with a collimated beam of atoms slowed down using a Zeeman slower. The slowed atoms are captured by a magneto-optical trap in the main experimental chamber. We usually load
$\sim$60 million atoms with temperature $\sim$70~$\mu$K into the magneto-optical trap over 10~s.
After loading we compress the atoms and use optical molasses to cool the atoms to $\sim$13~$\mu$K and transfer them to the internal ground state $F = 3$. The atoms are then transferred into the Raman lattice. 
With  degenerate Raman sideband cooling we cool about 20 million atoms in the absolute ground state $|F = 3, m_F = 3\rangle$ to 1~$\mu$K, which we then transfer into the large crossed dipole trap. 
After 0.5~s of plain evaporation in the dipole trap we start ramping up the dimple trap. The dimple beams are linearly ramped to their maximum powers in 1.5~s and held for another 400~ms to let the atoms thermalize. Then the large dipole trap is turned off and evaporation in the dimple trap begins.
The powers of dimple beams are exponentially ramped down in 6~s and after 2~s we start with additional exponential rampdown of the magnetic field gradient.
 The combination of the two techniques leads to efficient evaporation and at the end of it we produce a BEC of approximately 4000 atoms. More details can be found in Ref. \cite{Meznarsic2019} and \cite{Meznarsic2022}.

\subsection{Numerical simulation}

We use the split-step spectral method to solve the Gross-Pitaevskii equation 
\begin{equation}
 i \frac{\partial \psi}{\partial t} = - \frac{1}{2} \frac{\partial^2 \psi}{\partial x^2} + 2k(t)  \vert \psi \vert^2 \psi,
\end{equation}
in the dimensionless form \cite{PerezGarcia1998}, where time $t$ is given in units $1/\omega_r$, $\omega$ in units $\omega_r$, and distances in units $a_r = \sqrt{\hbar/m\omega_r}$. $k$ is the interaction parameter $k(t) = Na(t)/a_r$. We solve the equation on an uniform 1D grid of 65536 points in the interval [-1200:1200]. The initial state of the BEC is determined by an imaginary-time evolution, while
the calculation of the real-time dynamics is performed with a time-step of $\tau = $ 0.001. 
The code is written in Wolfram Language (Mathematica) and it can solve for the ground state through imaginary-time propagation of an initial approximation, and determine the
time-dependence in real-time.  Most of the computation time is spent in performing the (inverse) Fourier transform, which is performed highly efficiently in Mathematica, thus the performance of the code is comparable to that of compiled codes while it is significantly more flexible. Further details and examples can be found in Ref. \cite{ZitkoGithub}.

\subsection{Calibration of modulation amplitude}

We measured the amplitude of the modulation of interaction by measuring the splitting of a Feshbach resonance at 19.9~G \cite{ChinVuleticKermanEtAl2004}. In a typical measurement of Feshbach resonance, we measure the number of atoms remaining in the trap after switching the magnetic field to a chosen value for 100 ms. Around  the magnetic field corresponding to the Feshbach resonance we observe enhanced atom losses. We repeat the same measurement with the addition of an oscillating magnetic field at the various frequencies used for jet formation. In this case, we observe that the dip in the number of atoms is widened (see Supplemental Material Fig. S5). The width of the dip corresponds to the peak-to-peak amplitude of the magnetic field oscillations. For jet formation we modulate the magnetic field around 17.1~G, where the wide $s$-wave Feshbach resonance of cesium has the zero crossing \cite{ChinVuleticKermanEtAl2004}. The interaction as a function of the magnetic field around the zero crossing is known \cite{ChinVuleticKermanEtAl2004}, and we use it to determine the amplitude of the interaction modulation. By measuring the effect on the atoms, we directly measure the magnetic field at the position of the atoms. The measurements are done for the same amplitudes and frequencies as for jet formation, meaning that all the non-nonlinearities and other effect of the electrical circuit producing the oscillating magnetic field are taken into account, including the amplifier (Lab Gruppen IPD 1200).

\section{Acknowledgments}

We thank Cheng Chin and Matja{\v z} Gomil{\v s}ek for helpful discussions.
We would also like to thank Samo Beguš and Davorin Kotnik for their help with electronics.
This work was supported by the Slovenian Research Agency (research core Grants No. P1-0125, No. P1-0099 and P1-0416, and research projects No. J2-8191 and No. J2-2514).

\section{References}

\bibliographystyle{apsrev4-2-fmf-eng_1}
\bibliography{Bibliography}

\end{document}